\documentclass[runningheads]{llncs}
\usepackage{graphicx}

\newcommand{\sk}{\mathrm{sk}}
\newcommand{\pk}{\mathrm{pk}}
\usepackage{algorithm}
\usepackage{algpseudocode}
\usepackage{tikz}
\usetikzlibrary{arrows}
\usepackage{verbatim, color}

\usepackage{amssymb}
\usepackage{pifont}
\newcommand{\xmark}{\ding{55}}%

\usepackage[most]{tcolorbox}
\usepackage{varwidth}

\usepackage{algorithm,algpseudocode}
\usepackage{tcolorbox}

\usepackage{tikz}
\usetikzlibrary{matrix,shapes,arrows,positioning,chains, calc}
\usepackage{cleveref}

\usepackage{tikz}
\usetikzlibrary{arrows}
\usepackage{verbatim, color}
 \usepackage{mathtools}

\usepackage{cleveref}
\usepackage{msc}
\usepackage{algorithm,algpseudocode}
\usepackage{tcolorbox}

\usepackage{tikz}
\usetikzlibrary{matrix,shapes,arrows,positioning,chains, calc}

\begin{document}
\title{A Beyond-5G Authentication and Key Agreement Protocol\thanks{This work was supported by the Business Finland Consortium Project “Post-Quantum Cryptography” under Grant 754/31/2020.}}

\author{Mohamed Taoufiq Damir \and Tommi Meskanen\and \\ Sara Ramezanian \and
Valtteri Niemi}
\authorrunning{Damir. et al.}
%
\institute{Department of Computer Science, University of Helsinki\\ Helsinki Institute for Information Technology (HIIT), Helsinki, Finland. 
\\
\email{\{mohamed.damir,tommi.meskanen, sara.ramezanian,valtteri.niemi\}@helsinki.fi}}
%
%
%
%
\maketitle              
\begin{abstract}
The standardized Authentication and Key Agreement protocol for 5G networks (5G AKA) have several security and privacy vulnerabilities. In this paper, we propose a novel authentication and key agreement protocol for 5G and beyond that is compatible with the standardized 5G AKA. Our protocol has several privacy and security properties, e.g., perfect forward secrecy, resistance against linkability attacks, and protection against malicious SNs. Moreover, both the user identity protection and the perfect forward secrecy are handled using \textrm{Key Encapsulation Mechanisms} (KEM), which makes our protocol adaptable to the quantum-safe setting. To analyze the performance of the proposed protocol, we use the post-quantum KEM CRYSTALS-Kyber, recently chosen to be standardized by NIST, and NIST post-quantum Round 4 candidate KEMs. The results for communication and computation costs show that utilizing our protocol is feasible in practice and sometimes outperforms the public-key cryptography used in 5G AKA, i.e., ECIES. We further prove the security of our protocol by utilizing \textit{ProVerif}.
\end{abstract}

\section{Introduction}
The 5G technology positively impacts several industries, such as healthcare, transportation, and autonomous vehicles \cite{attaran2021impact}. However, the emergence of 5G has increased the concerns about the security and privacy of mobile users \cite{ahmad20175g}.
A proper authentication mechanism is essential to provide many services, e.g., roaming. The 3rd Generation Partnership Project (3GPP) group, responsible for the standardization of 3G, 4G, and 5G, specified the security architecture and procedures for 5G in its technical specification [\texttt{TS 33.501}]. A major component of the specification is the \textit{Authentication and Key Agreement} (AKA) protocol in 5G. Compared to previous mobile generations, the 5G specification emphasised more on user privacy, which is due to the importance and the high demand for privacy. We also recall that the 5G standards were developed while new privacy regulations have taken effect, for example, the European Union's General Data Protection Regulation (GDPR) \cite{regulation}. Unfortunately, and shortly after standardizing the 5G AKA by 3GPP, various security and privacy issues have been discovered, for example, linkability attacks  \cite{arapinis2012new,fouque2016achieving,borgaonkar2019new} and the lack of a full protection against compromised/impersonated SNs. The latter was identified during the 5G AKA Tamarin formal verification \cite{basin2018formal}. Moreover, passive/active attacks on 5G networks became a realistic threat \cite{chlosta20215g}, which is due to the gradually increasing availability of the necessary software and hardware to perform such (fake base station) attacks.

Among the new features in the 5G AKA, is the protection of the user identity using public-key cryptography, i.e., the ECIES algorithm. However, it is expected that such a solution would not last for a long period with the rise of quantum computers. A sufficiently large quantum computer can break many currently used cryptographic algorithms including the ECIES, see \cref{PQC}. Therefore, some of the mechanisms in 5G that are considered secure and private at the time of writing, such as users' identity protection, may be broken once large-scale quantum computers appear. It is unclear when such a quantum computer will be available, but it is worth mentioning that some leading companies such as Google and IBM are working on developing quantum computers and they are offering access to their computers over the cloud. Thus, the average attacker is expected to get access to quantum devices (over the cloud), while the average user is still using "classical" devices. Hence, there is interest in cryptography that works on classical devices but with the property of resisting quantum attacks, called \textit{Post-Quantum Cryptography} (PQC). 


In this context, we consider perfect forward secrecy, which is concerned with an attacker who is recording present (encrypted) sessions, with the hope of decrypting those at some future time point. The break would be possible if some long-term secrets were broken at that later time point, e.g., by using a quantum computer. 
 
In this work, we propose a novel AKA protocol that has several security and privacy properties, while keeping in mind further practical considerations such as backward compatibility with previous mobile generations and the adaptability to quantum-resistant cryptography. Our detailed contributions and the properties of our protocol are given in Section \ref{contributions}.

\section{Preliminaries}
\subsection{5G Terms and Acronyms}
For the purposes of our discussions, we reduce the mobile network architecture to three relevant entities: (1) The \textit{User Equipment} (UE) which further consists of the \textit{Universal Subscriber Identity Module} (USIM) and the \textit{Mobile Equipment} (ME), (2) the \textit{Home Network} (HN), and (3) the \textit{Serving Network} (SN). In 5G,  the Home Network assigns to every subscriber, and a USIM, a globally unique identifier that is called the \textit{Subscription Permanent Identifier} (SUPI). The SUPI can be used to track and locate users. For protecting user privacy, the SN assigns to the UE a \textit{Globally Unique Temporary Identity} (GUTI), which is a temporary and frequently changing identifier. The idea is to use, as often as possible, GUTI instead of SUPI but there are also circumstances where the GUTI cannot be used as an identifier. 

To avoid sending the SUPI as a plaintext, [\texttt{TS 33.501}] includes a mechanism to conceal the SUPI, resulting in \textit{Subscription Concealed Identifier} (SUCI). The SUPI concealment is done by using a HN public key, $\pk_H$, stored at the USIM with an \textit{Elliptic Curves Integrated Encryption Scheme} (ECIES). In 5G, the UE and the HN share a long term key K, and a sequence number SQN, where K is stored at the temper-resistant part of the USIM at the UE side, while SQN is used to check synchronization and detect replay attacks. The SUPI/GUTI, K and SQN are used to establish a mutual authentication and key agreement between the UE and SN via the HN.
\subsection{Key Encapsulation Mechanisms}
A \textit{key encapsulation mechanism} (KEM) is a scheme that is used in cryptographic protocols to exchange symmetric keys. A KEM is a triple of algorithms $\textbf{KEM}= (\textbf{KeyGen}(),\textbf{Encaps}(),\textbf{Decaps}())$ with a key space $\mathcal{K}$, where
\begin{itemize}
    \item $\textbf{KeyGen()}$ is a non-deterministic algorithm that generates a pair of public and secret key $(\pk,\sk)$.
    \item $\textbf{Encaps}(\pk)$ is a non-deterministic key encapsulation algorithm. The input of $\textbf{Encaps}(\pk)$ is $\pk$ and its outputs are a ciphertext $c$ and a key $k\in \mathcal{K}$.
    \item $\textbf{Decaps}(\sk,c)$ is a deterministic key decapsulation algorithm. The inputs of $\textbf{Decaps}(\sk,c)$ are $sk$ and $c$, and the algorithm returns a key $k\in \mathcal{K}$ or $\perp$ denoting failure.
\end{itemize}
The three algorithms work together in a natural fashion, e.g., both $\textbf{Encaps}$ and $\textbf{Decaps}$ produce the same key $k$ when the input keys $\pk$ and $\sk$ are chosen from the same pair. 
\subsection{Post-Quantum Cryptography}\label{PQC}
The security of the most popular practical public-key cryptography schemes is based on the hardness of integer factoring and the difficulty of calculating discrete logarithms. In \cite{shor1994}, Shor showed that these two problems can be efficiently solved by a sufficiently large-scale quantum computer.
In December 2016, the National Institute of Standards and Technology in the USA (NIST) initiated a standardization process by announcing a call for proposals for Post-Quantum (PQ) KEMs.
 In July 2022, NIST selected the CRYSTALS-Kyber KEM to be the primary algorithm recommended for standardization. Additionally, NIST picked four other algorithms for the next (4th) selection round. These four include BIKE, Classic McEliece, HQC, as well as the SIKE KEM. Our protocol uses "generic" KEMs, but for evaluating the protocol performance, we consider PQ KEMs. We give further details on implementing the mentioned algorithms\footnote{We discard the evaluation of SIKE as this algorithm has been shown to be insecure after it was selected.} as part of our protocol in Section \ref{practical}.

\subsection{Used Symmetric Primitives}\label{functions}
In order to be consistent with the 3GPP standardization, we use the same notations as in [\texttt{3GPP TS 33.501}]. In our protocol we use a \textit{Key Derivation Function} (KDF) which is based on SHA256, and seven symmetric key algorithms that are denoted by $f_1,f_2,f_3,f_4, f_5,f_1^*$ and $f_5^*$. Please note that although 3GPP did not fully standardize the above functions, 3GPP requires that breaking the security of these functions should require approximately $2^{128}$ operations. The MILENAGE algorithm set [\texttt{3GPP TS 35.205}] provides examples of the functions $f_1-f_5^*$ which utilize an AES-128 block cipher as a kernel function. Please note that the AES-256 is quantum-resistant and can be utilized as kernel function (because it has the same block size as AES-128).
\section{Related Work}\label{RW}
Various works have pointed out security and privacy issues in mobile network authentication and key agreement (AKA) protocols. One of these issues is related to linkability attacks. Such attacks consist of the attacker linking protocol executions based on the user's behavior to conclude some critical information about the user, for example, their identity or location. In \cite{arapinis2012new}, the authors described an attack where they exploited the failure messages in previous mobile AKA protocols to track the target user, and the authors proposed concealing the error messages using the HN public key. In \cite{fouque2016achieving}, Fouque et al. discovered another attack that accrues despite the fix proposed in \cite{arapinis2012new}. The work in \cite{borgaonkar2019new}, described a threat where the attacker can guess the pattern of the \textit{sequence number} by exploiting the synchronization failure message sent by a target UE. In our protocol, we abandon the use of sequence numbers to avoid potential de-synchronization attacks. 

The mentioned linkability attacks became a central issue in many recent works on 5G/6G AKA protocols, see for instance \cite{wang2021privacy} and the references therein. In \cite{basin2018formal}, the authors provided a formal verification of 5G AKA, where they pointed out further security issues in the studied protocol. They showed that if the SN is compromised, then the attacker can make the SN assign the session key to a different UE, that is because the session key and the user identifier are sent to the SN in two different messages. Moreover, the UE is unable to detect SN impersonation attacks before the key confirmation with the SN, which is not mandatory in 5G AKA. That is because there is no mechanism at the UE allowing this last to check if it is talking to the SN that has the identity that was verified by the HN. The last issue is due to the lack of a key confirmation message from the SN to the UE. 

Solutions to the SN related problems identified in \cite{basin2018formal} while considering linkability were studied in \cite{braeken2019novel,koutsos20195g}, but these works did not consider further security properties, e.g., perfect forward secrecy. 
In our quantum resistance context, we expect that an attacker is currently recording the (encrypted) messages sent between the UE and the SN in the hope of compromising the long-term keys of either the UE or HN by some other means, for example, a large-scale quantum computer. In the standardized 5G AKA and previous generations, compromising the long-term keys will imply compromising previous session keys. In other words, the property of perfect forward secrecy is not provided. Such an issue in mobile networks was studied in \cite{arkko2015usim,liu2021new,hojjati2020blockchain}. In both \cite{arkko2015usim,liu2021new}, the perfect forward secrecy is based on the intractability of the discrete logarithm problem. Consequently, these proposals are vulnerable to quantum attacks (Shor's algorithm). 
The work in \cite{hojjati2020blockchain} uses generic encryption in the protocol that we might assume to be post-quantum, but their protocol lacks protection against malicious and impersonated SNs. 

The perfect forward secrecy in \cite{arkko2015usim,liu2021new} follows from a Diffie-Hellman (DH) type key exchange. Thus, replacing DH in these works with a post-quantum key exchange would make them quantum resistant. However, at the time of writing no post-quantum key exchange is considered for standardization. In our work, both the SUPI protection and the perfect forward secrecy are based on KEMs which gives our protocol the possibility of implementing post-quantum KEMs in particular.  Implementing post-quantum KEMs in 5G was considered in \cite{ulitzsch2022post}, but the authors only consider the identification phase of the AKA protocol. Thus, further security features, e.g., linkability, and forward secrecy were not covered. 

As a side note on the work in \cite{liu2021new}, we would like to point out that the authors used ProVerif to formally verify their protocol. We remarked that the code published in \cite{liu2021new} considers the channel between the UE and SN as secure, while the channel between the SN and the HN is insecure. It should actually be the converse. We re-implemented the authors' verification with the corrected assumptions and noted that two of the claimed properties are then false (items 4 and 9, Section 5.1, pp. 324). We applied a minor correction to the Proverif code and item 9 turns out to be true after all. However, item 4 cannot be true. More precisely, item 4 states that the SN can distinguish a legit identification message (the first message in the protocol from the UE to SN) from an identity sent by an attacker, which is not ensured by the proposed protocol.
In the next section, we list our contributions in more detail.
\section{Contributions}\label{contributions}

We present a novel authentication and key agreement protocol for beyond-5G, where we consider various privacy and security issues from previous mobile generations. We summarise the security properties of our protocol as follows:

    (1) SUPI confidentiality;
    
    (2) Mutual authentication between UE and SN via the HN;
    
    (3) Confidentiality of the session key;
    
(4) Protection against the attacks in \cite{basin2018formal} (compromised/impersonated SNs);

(5) Perfect forward and backward secrecy;

(6) Unlinkability and protection against replay attacks;

(7) Compatibility with the KEM paradigm, esp. with standardized PQ KEMs.

Moreover, we use the formal verification tool ProVerif to prove some of the above claims. Furthermore, we give an overview of practical implementation of our protocol. First, we describe some backward compatibility properties of our protocol with previous mobile generations, see Section \ref{compatibility}. Second, we discuss implementing the protocol using  Kyber, the post-quantum KEM recently selected by NIST and round4 KEM NIST finalists, see Section \ref{practical}. We show that implementing our protocol with such KEMs, and especially Kyber, outperforms the public-key cryptography used nowadays (i.e., ECIES) in 5G AKA.

Table \ref{comparison with other works} compares the security and privacy properties of our protocol with those properties of several recent works.

 \section{Our Protocol}
The proposed protocol consists of two phases. Phase A is the identification phase, where the UE is identified by the HN. Phase B is an authentication phase, which allows the UE and the HN to securely authenticate each other.

 \subsection{Phase A: The Identification Phase}
 The identification phase consists of three steps, a message from the UE to the SN, a message from the SN the the HN, and an identification confirmation/abortion at the HN.
 
\begin{table}[t]
\begin{center}
\resizebox{0.70\columnwidth}{!}{%
\begin{tabular}{|l|l|l|l|l|l|l|l|l|l|l|l|}
\hline
Properties &  \cite{arkko2015usim} &  \cite{liu2021new} &  \cite{braeken2019novel} &\cite{koutsos20195g} &5G AKA & \cite{wang2021privacy} & \cite{hojjati2020blockchain} & Ours \\ \hline
Unlinkability &   \xmark    &  \checkmark  &  \checkmark     &  \checkmark   &   \xmark   &   \checkmark  &  \checkmark&\checkmark     \\ \hline
Perfect Forward Secrecy   &  \checkmark    &   \checkmark   &   \xmark   &  \xmark  &   \xmark   &   \xmark   &  \checkmark & \checkmark    \\ \hline
Compromised/impersonated SN protection      &    \xmark   &   \checkmark    &    \checkmark   &   \checkmark  &    \xmark  &   \xmark   &   \checkmark &\checkmark  \\ \hline
Quantum-Safe       &   \xmark   &  \xmark    &   \xmark   &   \xmark  &  \xmark    &  \xmark    &  \checkmark & \checkmark  \\ \hline
Coverage over GUTI     &   \checkmark   &  \xmark    &   \xmark   &   \checkmark &  \checkmark    &  \xmark    &    \xmark &\checkmark  \\ \hline
\end{tabular}}
\end{center}
\caption{Comparison of the security and privacy properties of our protocol with the prior art.}
\label{comparison with other works}
\end{table}
In our protocol, and as in 5G AKA, the parameters at the UE are the SN and HN identities denoted respectively by $ID_{SN}$ and $ID_{HN}$. The USIM stores also, $K$, a shared long term key with the HN, and the HN public key $pk_H$. In our protocol, $pk_H$ is a KEM key.
The communication between the UE and SN is either initiated by the UE, e.g., for outgoing call, or by the SN, e.g., for incoming call. In both cases, the UE has to send an identifier (SUCI or GUTI) to the SN. In our context, the two cases are similar. In the rest of this section, we cover the SUCI case, while the GUTI case is covered in Section \ref{GUTI}. The identification procedure for the SUCI case goes as follows:

\begin{figure}[]
\centering
\begin{minipage}{0.53\textwidth}
\scalebox{.69}{
\begin{tcolorbox}[title={ {\tt Identification\_Response}\\ ($ \textrm{SUPI}, , \textrm{ID}_{SN}, \textrm{ID}_{HN}, pk_H$)},center title,hbox]
       \begin{varwidth}{\textwidth}

 \begin{enumerate}
    
     \item At ME $(\pk_U,\sk_U)\leftarrow \textbf{KeyGen()}$.
     \item $(c_1,K_{s_1})\leftarrow \textbf{Encaps}(\pk_{HN}).$
     \item $\textrm{SUCI}_{conc}\leftarrow \textrm{Enc}_{K_{s_1}}(\textrm{SUPI} || \pk_U, \textrm{ID}_{SN})$.
     \item $\textrm{MAC}_U=HMAC(\textrm{SUCI}_{conc}, K_{s_1})$.
     \item $\textbf{Send } (c_1, \textrm{SUCI}_{conc},\textrm{MAC}_U, \textrm{ID}_{HN} )$ to SN.
 \end{enumerate}
 \end{varwidth}
\end{tcolorbox}}
\caption{Identification response \\ from the UE to HN}\label{idrequest}
 \end{minipage}
\begin{minipage}{0.46\textwidth}
\scalebox{.73}{
\begin{tcolorbox}[title={
${\tt Identification\_at\_HN}$\\ ($\sk_{H}$, $\textrm{ID}_{SN}, c_1,  \textrm{SUCI}_{conc},\textrm{MAC}_U$)},center title,hbox]
       \begin{varwidth}{1.18\textwidth} 

 \begin{enumerate}
     \item $K_{s_1}\leftarrow \textbf{Decaps}(c_1, \sk_{HN}$).
     \item $(\textrm{SUPI}, \pk_U,\textrm{ID}_{SN})\leftarrow \textrm{Dec}_{K_{s_1}}(\textrm{SUCI}_{conc})$.
     \item $\textrm{MAC}_U'=HMAC(\textrm{SUCI}_{conc}, K_{s_1})$.
     \item \textbf{Check} whether $\textrm{MAC}_U'=\textrm{MAC}_U$ and verify $\textrm{ID}_{SN}$; abort in the negative case.
    
 \end{enumerate}
 \end{varwidth}

       \end{tcolorbox}}\caption{Identification of the UE at the HN}\label{idHN}
 \end{minipage}
\end{figure} 
\begin{enumerate}
  \item The SN sends an identification request to UE.
\item Identification response from the UE (Figure \ref{idrequest}):
\begin{itemize}

    \item The ME starts by freshly generating a pair of public/private KEM keys $(\pk_U,\sk_U)$.
    \item The ME generates a shared KEM key $K_{s_1}$ and a cipher text $c_1$ using an encapsulation algorithm and the HN public key $pk_H$.
    \item  The ME conceals the SUPI, $\pk_U$ and $ID_{SN}$ using the key $K_{s_1}$ and a symmetric encryption $\textrm{Enc}_{K_{s_1}}$. We denote such a concealment by $\textrm{SUCI}_{conc}$.
    \item The ME computes a MAC tag using HMAC with $\textrm{SUCI}_{conc}, K_{s_1}$.
    \item Finally, the UE sends $(c_1, \textrm{SUCI}_{conc},\textrm{MAC}_U, \textrm{ID}_{HN} )$ to SN.
\end{itemize}

\item The message from SN to HN: 
Once the SN receives $c_1, \textrm{SUCI}_{conc},\textrm{MAC}_U$, and $ \textrm{ID}_{HN}$ from UE, the SN generates a random 256 bit string $R_{SN}$, and then forwards $(c_1, \textrm{SUCI}_{conc},\textrm{MAC}_U, R_{SN} )$ to the HN.

\item The UE Identification at the HN (Figure \ref{idHN}): 
The parameters at the HN are, $sk_H$, the KEM secret key (bound to $\pk_H$), and $ID_{SN}$.
Once the HN recieves  $(c_1, \textrm{SUCI}_{conc},\textrm{MAC}_U, R_{SN} )$ from the SN, the HN proceeds on identifying the UE using the algorithm in Figure \ref{idHN}, where $\textrm{Dec}_{K_{s_1}}$ denotes a symmetric decryption using the key $K_{s_1}$, which is resulting from the decapsulation $\textbf{Decaps}(c_1, \sk_{HN})$. 
\end{enumerate}

The next step consists of the generation of an authentication challenge for the UE by the HN.  
\subsection{Phase B: The Authentication Phase}\label{auth}
Once the MAC check passes at the HN (Figure \ref{idHN}), the HN retrieves the UE's long term key K based on SUPI, derives the key $K_{s_2}$ and the ciphertext $c_2$ using the appropriate KEM encapsulation algorithm and $\pk_U$. In more detail, we have:
\begin{enumerate}
    \item The HN computes an authentication vector using  
    the  ${\tt Auth\_Vector}$ algorithm depicted in Figure \ref{AV}. 

\begin{figure}[]
\centering
\begin{minipage}{0.52\textwidth}
\scalebox{.73}{
\begin{tcolorbox}[title={
 ${\tt Auth\_Vector}$($K, \pk_{U}, K_{s_2}$, $R_{SN}, \textrm{ID}_{SN}$)},center title,hbox]
       \begin{varwidth}{\textwidth} 
 \begin{enumerate}
\item $(c_2,K_{s_2})\leftarrow$ \textbf{Encaps}($\pk_{U}$).
     \item Compute $MAC=f_1(K,K_{s_2},R_{SN})$.
     
     \item XRES=$f_2(K,K_{s_2})$.
     \item $CONC=f_5(K,K_{s_2})\oplus R_{SN}$.
     \item AUTN=$\textrm{CONC}||\textrm{MAC}$.
    \item CK=$f_3(K,K_{s_2})$, IK=$f_4(K,K_{s_2})$
     \item $\textrm{XRES}^*$=$\textrm{KDF(CK,IK},K_{s_2},\textrm{XRES},ID_{SN})$.
     \item $\textrm{HXRES}^*$=$SHA256(R_{SN},\textrm{XRES}^*)$
\item  $K_{ausf}=\\ \textrm{KDF}(CK,IK,K_{s_2},CONC,ID_{SN})$.
    \item  $K_{seaf}=\\ \textrm{KDF}(K_{ausf},ID_{SN})$.
    \item Set $K_3=\textrm{XRES}^*\oplus f_5(K,K_{s_2})$.
    \item Compute the encryption \\ $M=\textrm{SEnc}_{K_3}(K_{seaf}, \textrm{SUPI})$ 
     \item \textbf{Return} (AUTN, $\textrm{HXRES}^*$, M)
 \end{enumerate}
     \end{varwidth}

       \end{tcolorbox}}
 \caption{Challenge and key material generation at HN\label{AV}}
\end{minipage}
\begin{minipage}{0.47\textwidth}
\scalebox{.73}{
\begin{tcolorbox}[title={{\tt USIM\_Resp}($K_{s_2}, R_{SN}$, AUTN, K)},center title,hbox]
       \begin{varwidth}{\textwidth}
     \begin{enumerate}
 \item Parse AUTN.
\item Compute $ AK=f_5(K,K_{s_2})$
\item Compute $R_{SN}$ from $CONC=AK\oplus R_{SN}$
\item Check $f_1(K,K_{s_2},R_{SN}) =\textrm{MAC}$. 
\item If check does not pass, abort.
\item $\textrm{RES}=f_2(K,K_{s_2})$.
\item CK=$f_3(K,K_{s_2})$, IK=$f_4(K,K_{s_2})$.
\item \textbf{Return} ($\textrm{RES}$, CK, IK)
 \end{enumerate}
       \end{varwidth}

       \end{tcolorbox}}
\caption{Challenge response at the USIM  \quad\quad \label{atUSIM}}

\scalebox{.75}{

\begin{tcolorbox}[title={{\tt ME\_Resp}($K_{s_2}$, CK, IK, AUTN, $\textrm{ID}_{SN}$, $\textrm{RES}$)},center title,hbox]
       \begin{varwidth}{\textwidth}
      \begin{enumerate}
 \item $\textrm{RES}^*=\\ \textrm{KDF(CK,IK,} K_{s_2}, \textrm{RES},ID_{SN})$.
 \item Get $CONC$ from AUTN.
\item  $K_{ausf}=\\ \textrm{KDF(CK,IK,} K_{s_2}, \textrm{CONC},\textrm{ID}_{SN}).$
\item  $K_{seaf}=\textrm{KDF}(K_{ausf},ID_{SN})$.
\item \textbf{Return} ($K_{seaf}, \textrm{RES}^*$)
 \end{enumerate}
\end{varwidth}
\end{tcolorbox}}
\caption{Session key and $\textrm{RES}^*$ generation \\ at the ME.}\label{atME}
\end{minipage}
\end{figure}

\item Next, the HN sends AUTN, M, $\textrm{HXRES}^*$ and $c_2$ to the SN.
\item The SN forwards both AUTN and $c_2$ to the UE.
\item At the UE side:
\begin{itemize}
    \item The ME first obtains the key $K_{s_2}$ using the decapsulation algorithm with the stored secret key $\sk_U$, namely
$K_{s_2}\leftarrow \textbf{Decaps}(c_2, \sk_U)$.
\item The USIM receives $K_{s_2}$ from the ME, then proceeds on computing the response to the challenge using the ${\tt USIM\_Resp}$ algorithm in Figure \ref{atUSIM}.
\item The ME generates $\textrm{RES}^*$ and the key material using ${\tt ME\_Resp}$ as depicted in Figure \ref{atME}.
\item The ME forwards $\textrm{RES}^*$ the SN.
\end{itemize}

\item The SN receives the value of $\textrm{RES}^*$ from the UE and then
\begin{itemize}
    \item compares $\textrm{SHA256}(\textrm{RES}^*, R_{SN})$ with $\textrm{HXRES}^*$. If the two values are not equal, then abort.
    \item the SN uses $R_{ SN}$ to obtain $f_5(K,K_{s_2})$ from CONC.
    \item computes $K_3=\textrm{RES}^*\oplus f_5(K,K_{s_2}).$
    \item The SN obtains the SUPI and $K_{seaf}$ by decrypting $M$ using $K_3.$
\item The SN sends a confirmation message to the HN.

\end{itemize}

\end{enumerate}
\subsection{Remarks on the Enhancements Required by Our Protocol}\label{compatibility} 
In mobile telephony, and from a design point of view, it is desirable to implement novel protocols ensuring backward compatibility with previous mobile generations.  In our protocol, the authentication response at the UE uses the same algorithms as in 5G AKA, see \texttt{TS 33.501}, namely, {\tt ME\_Resp} and {\tt USIM\_Resp} depicted in Figures \ref{atME} and \ref{atUSIM}. The only difference between our proposed protocol and the UE computations in 5G AKA is the used input for such algorithms. More precisely, we use $R_{SN}$ instead of the \textit{sequence number} (SQN) and we use $K_{s_2}$ instead of a random bitstring generated at the HN. Consequently, our protocol would require only minor extensions at the USIM level.

\section{The Case of GUTI}\label{GUTI}

In the above, we covered the case where the identification is based on the permanent identifier SUPI. However, the much more frequent case is when the \textit{Global Unique Temporary Identifier }(GUTI) is used. The use of GUTI is favored over the use of SUPI in 5G (and also in earlier generations) because it is a frequently changing identifier that is chosen independently of SUPI. When GUTI is used as an identifier, no (asymmetric) encryption is required at the UE side. In the best case, after every successful communication, the SN assigns a new GUTI to the UE over the established secure channel. 

As in the case of SUPI, the communication can be initiated either by the UE or the SN. In the rest of this section, we assume that the UE identification is based on GUTI, where the UE sends its GUTI to the SN. Next, the SN either chooses to continue using the shared session keys from the previous connection or starts a fresh authentication via the HN. Note that in both cases a new GUTI can be allocated to the UE. In the latter case, the authentication (in 5G AKA) is similar to the SUPI case. Hence, the resulting session key $K_{seaf}$ will suffer from the mentioned security issues. In this section, we equip our protocol with a mechanism covering the GUTI case.

We recall that after a successful run of our protocol, the UE and the HN share the key $K_{s_2}$ and the random bit string $R_{SN}$ generated by the SN. We further require that after each successful protocol run, both the UE and the HN store a hash value $$K_S=h(K_{s_2}, R_{SN}),$$ where $h$ is an appropriate standard hash function. After storing $K_S$, both $K_{s_2}$ and $R_{SN}$ are deleted.
Moreover, we require that with every GUTI assignment, the SN generates and sends, in addition to the GUTI, a random bitstring $R_{SN}'$ to the UE over the established secure channel. The idea behind our solution for the GUTI case is to replace $K_{s_2}$ with $$K_{S}'=K_S\oplus R_{SN}'.$$ in our SUPI-based protocol. The procedure goes as follows:
\begin{enumerate}
    \item In the beginning of the connection establishment, the UE identifies itself by sending GUTI.
    \item The SN resolves the SUPI from GUTI and forwards the SUPI with the stored $R_{SN}'$ and a freshly generated new random bitstring $R_{SN}$ to the HN.
    \item The HN notices that GUTI has been used as an identifier, computes $K_{S}'$ based on $R_{SN}'$ and the stored $K_{S}'$, and runs $\texttt{Auth\_Vector}$ with $K_{S}'$ instead of $K_{s_2}$.
    \item Similarly, the UE runs $\texttt{AT\_ME}$ and $\texttt{USIM\_Resp}$ with $K_{S}'$ instead of $K_{s_2} $.
    \item After all steps in the protocol have been completed succesfully, both the UE and the HN  replace $K_S$ with $h(K_{S}', R_{SN})$. Then everything is ready for another run of a GUTI-based authentication.
    
\end{enumerate}

In the last step above, the HN may delete the old $K_S$ because it knows that the UE has completed the protocol successfully. On the other hand, the UE has to keep also the old $K_S$ until it gets confirmation from the SN about successful completion of the whole protocol also on the HN side. This confirmation may be given in several ways, either explicitly or implicitly, but we leave the details out of scope of this paper.

Please note that the forward security in the case of SUPI (resp., GUTI) is based on the shared $K_{s_2}$ (resp., $K_S$), while the protection against compromised/impersonated SNs follows from the contribution of the SN, i.e., $R_{SN}$ and the MAC check at the UE in both the SUPI and the GUTI case. By assumption, the parameters $K_{s_2}, K_S$ are shared only between the UE and the HN, while $R_{SN}$ and $R_{SN}'$ are shared by the UE, SN and HN. Moreover, the SUPI and GUTI protocols are similar, and the only difference consists of replacing $K_{s_2}$ by $K_S$.  Furthermore, and thanks to the hash function $h$, it is practically impossible to link $K_{s_2}$ to $K_S$. The same is true for $R_{SN}$ and $R_{SN}'$ as they are randomly and independently generated. Consequently, a SUPI based protocol execution and a subsequent GUTI based protocol execution cannot be linked to each other.
Due to the similarity between the GUTI and SUPI cases, we mainly focus on the security analysis of the SUPI based protocol.
\section{Security Analysis}
We prove the security of our protocol by utilizing ProVerif \cite{blanchet2008automated}, which is one of the well-known formal verification tools.
\subsection{Threat Model}
We assume a Dolev-Yao model for attackers. Moreover, to evaluate the forward-secrecy property, we consider a more powerful attacker, i.e., an attacker that can compromise some parties. Thus, we additionally consider the eCK model \cite{eCK}. Our formal verification can be split into two parts. The first part consists of proving the security of a \textit{clean session}. During this session, the adversary cannot control any party and does not have access to the parties' long-term or temporary keys. More precisely, we prove that after the execution of a clean session, our protocol ensures authentication and secrecy of the SUPI, long-term key $K, \sk_{HN}$, and the session key. In the second part of our verification, we assume that the  HN or UE (or both) are compromised; we then prove that our protocol satisfies the forward-secrecy property under such assumptions.

Next, we precise our assumptions on the protocol's channels and components. Our assumptions are drawn from the 5G requirements specified in \texttt{TS 33.501}.

\textbf{Assumptions on the Channels}
As in the case of 5G, our protocol uses two separate channels. The first one is the radio channel between the UE and the SN; see the 5G specification, \texttt{TS 33.501}. We assume the presence of a Dolev-Yao attacker who can intercept, manipulate and replay messages on this channel. The second channel is a wired channel between the SN and HN; in contrast with the above radio channel, the channel between the SN and HN is explicitly specified by \texttt{TS 33.501} as a e2e core network interconnection channel. Consequently, we adopt the assumption that such a channel is secure, namely, a channel that provides both confidentiality and integrity. 

\textbf{Assumptions on the Components}
We recall that our protocol consists of three components, the UE, the SN and the HN. Our assumptions on the protocol components are the following: (1) The UE consists of the USIM and the ME. In our protocol we assume that both the asymmetric (post-quantum) encryption and the session key $K_{seaf}$ derivation are performed by the ME, where the ME uses parameters that are given by the USIM. In our model, we consider the UE to be one single secure entity. More precisely, the exchange between the ME and USIM is assumed to be secure and the key $K_{seaf}$ is protected at the UE after the execution of the protocol. Moreover, the long term key $K$ is residing at the temper-resistant part of the USIM, thus, assumed to remain protected. (2) The attacker cannot obtain the key $K_{seaf}$ at the SN. (3) The long term parameters at the HN, i.e., $K$ and $\sk_{H}$ are protected during the protocol execution. Note that in the context of forward secrecy we assume that such parameters leaked after an honest execution of the protocol.  
\subsection{Formal Verification}\label{FV}
Our verification consists of four parts. The process at the UE, the process at the SN, the process at the HN, denoted by \texttt{UE, SN, HN} respectively, and a main process to conclude to proof. Our ProVerif code with implementation details and design choices on the chosen primitives, i.e., XOR, KEM. is available in our repository at  \url{https://github.com/Secure-6G/ProVerif-AKA-6G}.

\textbf{Verification Results}
Our verification shows  the excutability of our protocol by showing that each pair of successive messages is executed in sequence. We further prove the secrecy of the protocol long term parameters, namely, $K$ the long term key at the UE/HN, $\sk_{HN}$, the secret key at the HN and the long term identifier SUPI. Moreover, the authentication of the UE by the SN by the help of the HN is proved. Finally, Proverif shows that forward-secrecy holds even if the long term keys at the UE and HN have been compromised. 

\subsection{Further Security Features}

\textbf{Protection against compromised and impersonated SNs}\label{confirmation}
As mentioned in Section \ref{RW}, 5G AKA suffers from two attacks identified in \cite{basin2018formal}. The first attack follows from an attacker compromising the SN and resulting in making the SN to assign the wrong SUPI to the session key. As pointed in \cite{basin2018formal}, this attack is due to the fact that the SUPI and the session key are sent in two different messages (and not bound together). The second attack results from the lack of a mechanism allowing the UE to detect if the received authentication is sent by correct SN. 

In our protocol, both the session key and the SUPI are sent to the SN as a single encrypted message $M$ during authentication. 
At this stage, the SN is not able to obtain the key $K_3$ used to encrypt $M$, as it depends on $\textrm{XRES}^*$, the expected challenge response from the UE. In fact, the SN will only receive the response from the UE after the latter has authenticating both the HN and the SN, thanks to the MAC check at the UE, which depends on $R_{SN}$ and $K$.
Consequently, both binding the SUPI and the session key to each other, and checking the validity of the message sent by the SN to the UE are achieved by the above procedure.

\textbf{Preventing Linkability and Replay Attacks}
A linkability attack is an attack where the attacker is able to distinguish if different sessions belong to the same user. Examples of linkability attacks on mobile networks protocols include failure message linkability attack \cite{arapinis2012new}, encrypted SUPI replay attack \cite{fouque2016achieving} and sequence number inference attack \cite{borgaonkar2019new}, which are all particular cases of replay attacks. Our protocol prevents such attacks by providing resistance to replay attacks. Assuming that the channel between the SN and the HN is secure, then the vulnerable message flow of our protocol reduces to the messages sent between the UE and the SN. There is only one message from the SN to the UE consisting of the ciphertext $c_2$ resulting from an encapsulating using the freshly generated UE public key and the quantity $\textrm{AUTN}$  consisting of $f_5(K,K_{s_2})\oplus R_{SN}$  and $\textrm{MAC}=f_5(K,K_{s_2},R_{SN}).$
Assume that the ciphertext $c_2$ is produced by a KEM ensuring the Indistinguishability under Chosen Plaintext Attack (IND-CPA), which is the case for the post-quantum KEMs used in the next section to evaluate our protocol. Then any replayed cipher $c_2$ in our protocol will be detected by the UE. Moreover, the MAC in AUTN depends on the freshly generated key $K_{s_2}$ and randomness $R_{SN}$. We recall that $K_{s_2}$ is protected by the freshly generated secret key $\sk_U$ of the UE, while $R_{SN}$ is obtained using K and $K_{s_2}$. Consequently, replayed, handcrafted $c_2$, AUTN (or both) will be detected by the UE via the mentioned MAC check.

\textbf{Post-Quantum Forward Secrecy}
A general technique to achieve forward secrecy is to equip every protocol execution with a "fresh" authenticated Diffie-Hellman (DH) key exchange, that is to create an independence between the session key and the long term keys. This approach was used, e.g., in mobile telephony AKA protocols in \cite{arkko2015usim,liu2021new}. In theory, such proposals can be modified to become quantum secure by replacing DH by a similar post-quantum key exchange. At the time of writing, PQ key exchanges are still not at the same maturity level as KEMs. Moreover, using KEMs instead of DH have been shown to be more efficient in some contexts, e.g., the TLS protocol \cite{KEMTLS}. Thus, we argue that in this direction, a novel feature of our protocol is in the use of the (post-quantum) KEM paradigm to ensure perfect forward secrecy in 5G and beyond instead of using DH type key exchange. 

          \section{Feasibility of Our Protocol}\label{practical}
        In this section, we discuss potential implementations of our protocol. Our aim is to compare our protocol with the standardized 5G AKA. At the UE side (resp., HN side), we use the same symmetric primitives as in 5G AKA, with one hash extra hash $f_1(\pk_U,K)$ (resp., one extra hash and one symmetric encryption). Hence, we will restrict our comparison to asymmetric primitives, i.e., KEMs. For the SN, our protocol is only required to perform a random number generation, a symmetric decryption and an XOR operation, as additional operations compared to 5G AKA, thus we only focus on the operations at the UE and HN.

\textbf{Computational cost}
          At the UE side, our protocol uses one KeyGen and one Encaps before sending the first message to the SN, then the UE decapsulates the ciphertext $c_2$ sent by the HN.
          In 5G AKA, the UE uses one ECIES KEM encapsulation for the SUPI encryption, Hence, comparing the running time of asymmetric primitives in our protocol (at the UE) with 5G AKA reduces to comparing the time  of one ECIES key encapsulation with the running time of a KEM's $\textbf{KeyGen}+\textbf{Encaps}+\textbf{Decaps}.$ For the SUPI encryption in 5G AKA, \texttt{TS 33.501}, gives the options of using two ECIES profiles, namely, with Curve25519 and Secp256r1. For our protocol, we use CRYSTALS-Kyber recently chosen by NIST for standardization, and the potential algorithms for further standardization: BIKE, Classic McEliece, and HQC. Moreover, and to fairly compare the mentioned algorithms, we use the KEM implementations with parameters offering the security level required by 3GPP, which is equivalent to 128-bits AES. The computational cost figures of the asymmetric primitives at the UE are depicted in Table \ref{run}.
 Table \ref{run} is generated using liboqs \cite{liboqs}, which is an open source C library for quantum-safe cryptographic algorithms, where the implementations are directly from NIST submissions. For the ECIES profiles we use OpenSSL. The computations are performed on a 3.5 GHz Core (i7-7567U).
 
 We recall that our aim is to rank the implementations when various algorithms are used. Hence, the used device is not relevant in this context (mobile phone or workstation).
Using Table \ref{run}, we remark that Kyber outperforms both ECIES profiles. The most significant computational cost comes from Classic McEliece. It is worth mentioning that the heaviest operation in Classic McEliece is the KeyGen which takes approximately 14 ms. 

Next, we evaluate the operations at the HN using an approach similar to the operations at the UE. Compared to 5G AKA, which requires only one ECIES decapsulation at the HN, our protocol uses one Decaps and one Encaps at the HN (plus one hash and one symmetric encryption). 
Similarly to the case of UE, Kyber outperforms both ECIES profiles, similarly to the computations at the UE, the use of Classic McEliece at the HN is faster than ECIES Secp256r1.

\textbf{Communication cost}
Table \ref{communication} illustrates the (KEM) parameter sizes of the used schemes. 
In 5G AKA, the UE sends a SUCI, a MAC and a KEM ciphertext over the radio channel to the SN and this last forwards these parameter to the HN. Next, the HN forwards an authentication vector to the UE via the SN. Finally, the UE sends a challenge response to the HN via the SN. In our protocol, and in addition to the 5G AKA messages, the UE sends a symmetric encryption of a KEM public key, while the HN responds with an additional KEM ciphertext to the UE. From Table \ref{communication} we clearly see that ECIES provides a small communication cost compared to post-quantum KEMs, while Classic McEliece has the most significant communication cost due to the size of the public key. We note that the standardized Kyber offers the best communication cost among the studied PQ KEMs. Please note that our comparison considered that some quantum resistant scheme must replace ECIES in the near future. Hence, the focus is on determining which PQ KEM is most suitable for our scheme. 
          
\begin{table}[t]
\begin{minipage}{0.45\textwidth}
\begin{center}
\begin{tabular}{|l|l|l|}
\hline
Algorithm                                                                       & At UE & At HN \\ \hline
\begin{tabular}[c]{@{}l@{}}ECIES Curve25519 \end{tabular}             & 0.040  & 0.040      \\ \hline
\begin{tabular}[c]{@{}l@{}}ECIES Secp256r1 \end{tabular}              & 0.180  & 0.180       \\ \hline
\begin{tabular}[c]{@{}l@{}}Kyber \end{tabular}           & 0.026  & 0.019   \\ \hline
\begin{tabular}[c]{@{}l@{}}Classic McEliece \end{tabular} & 14.047 &0.047    \\ \hline
\begin{tabular}[c]{@{}l@{}}BIKE \end{tabular}            & 0.882  &0.672   \\ \hline
\begin{tabular}[c]{@{}l@{}}HQC \end{tabular}             & 0.421  &0.331  \\ \hline
\end{tabular}
\end{center}
\caption{Running time of the\\ asymmetric primitives at the UE \\and HN (milliseconds).}
\label{run}
\end{minipage}
\begin{minipage}{0.50\textwidth}
\begin{center}
\begin{tabular}{|l|l|l|l|l|}
\hline
Algorithm & $\sk$  & $\pk$ & Cipher &  key \\ \hline
ECIES Curve25519        & $32$      & $32$      & $32$      & $32$          \\ \hline
ECIES Secp256r1       & $32$      & $32$      & $32$      & $32$          \\ \hline
Kyber               & $1632$      & $800$      & $768$      & $32$          \\ \hline
Classic McEliece & $6452$      & $261120$   & $128$      & $32$          \\ \hline
BIKE         & $5223$      & $1541$      & $1573$      & $32$          \\ \hline
HQC     & $2289$       & $2249$      & $4481$      & $64$          \\ \hline

\end{tabular}
\end{center}
\caption{ Communication cost (bytes)}
\label{communication}
\end{minipage}
\end{table}
\section{Conclusion}
We presented an authentication and key agreement protocol for 5G and beyond with further security and privacy features that are not offered by the standardized 5G AKA. Such features include resistance to known linkability attacks, perfect forward secrecy, and protection against compromising and impersonating the SN. Moreover, in our protocol, we abandoned the use of sequence numbers to avoid possible desynchronization attacks. Furthermore, our protocol covers the case of GUTI which is usually ignored in similar works. We used Proverif to formally verify some of our security claims. Finally, we gave an overview of potential  implementations of the protocol using NIST post-quantum KEM candidates.  In summary, we illustrated a theoretical and practical implementation of a (quantum) safe AKA protocol for beyond-5G and presented a supporting argument for its security features using both formal and classical methods. As mentioned in the threat model, active attackers between the USIM and ME are omitted. While not considered in our protocol, we emphasize that our protocol does not prevent an attacker from requesting the session key from the USIM or ME, especially an attacker able to compromise the long term key K. This problem is left as future work.

%
%

\begin{thebibliography}{8}
\bibitem{shor1994}Shor, P. Algorithms for quantum computation: discrete logarithms and factoring. {\em Proc. 35th Annual Symposium FOCS}. pp. 124-134 (1994)

\bibitem{wang2021privacy}Wang, Y., Zhang, Z. \& Xie, Y. {Privacy-Preserving} and {Standard-Compatible}{AKA} Protocol for 5G. {\em 30th USENIX Security Symposium (USENIX Security 21)}. pp. 3595-3612 (2021)
\bibitem{basin2018formal}Basin, D., Dreier, J., Hirschi, L., Radomirovic, S., Sasse, R. \& Stettler, V. A formal analysis of 5G authentication. {\em Proceedings Of The 2018 ACM SIGSAC Conference On Computer And Communications Security}. pp. 1383-1396 (2018)
\bibitem{braeken2019novel}
Braeken, A., Liyanage, M., Kumar, P. \& Murphy, J. Novel 5G authentication protocol to improve the resistance against active attacks and malicious serving networks. 
{\em Ieee Access}.
 \textbf{7} pp. 64040--64052 (2019)
\bibitem{koutsos20195g}Koutsos, A. The 5G-AKA authentication protocol privacy. {\em 2019 IEEE European Symposium On Security And Privacy (EuroS\& P)}. pp. 464-479 (2019)
\bibitem{arapinis2012new}Arapinis, M., Mancini, L., Ritter, E., Ryan, M., Golde, N., Redon, K. \& Borgaonkar, R. New privacy issues in mobile telephony: fix and verification. {\em Proceedings Of The 2012 ACM CCS}. pp. 205-216 (2012) 
\bibitem{fouque2016achieving}
Fouque, P., Onete, C. \& Richard, B. Achieving Better Privacy for the 3GPP AKA Protocol.. 
{\em Proc. Priv. Enhancing Technol.}.
 \textbf{2016}, 255--275 (201)
\bibitem{borgaonkar2019new}
Borgaonkar, R., Hirschi, L., Park, S. \& Shaik, A. New privacy threat on 3G, 4G, and upcoming 5G AKA protocols. 
{\em Proc. Priv. Enhancing Technol.}.
 \textbf{2019}, 108--127 (201)
 






\bibitem{liboqs}Liboqs liboqs.  (2019), https://github.com/open-quantum-safe/liboqs

\bibitem{liu2021new}Liu, T., Wu, F., Li, X. \& Chen, C. A new authentication and key agreement protocol for 5G wireless networks. {\em Telecommunication Systems}. \textbf{78}, 317-329 (2021)
\bibitem{arkko2015usim}Arkko, J., Norrman, K., Näslund, M. \& Sahlin, B. A USIM compatible 5G AKA protocol with perfect forward secrecy. {\em 2015 IEEE Trustcom/BigDataSE/ISPA}. \textbf{1} pp. 1205-1209 (2015)
\bibitem{attaran2021impact}Attaran, M. The impact of 5G on the evolution of intelligent automation and industry digitization. {\em J Ambient Intell Humaniz Comput}. pp. 1-17 (2021)
\bibitem{ahmad20175g}Ahmad, I., Kumar, T., Liyanage, M., Okwuibe, J., Ylianttila, M. \& Gurtov, A. 5G security: Analysis of threats and solutions. {\em 2017 IEEE Conference On Standards For Communications And Networking (CSCN)}. pp. 193-199 (2017)


\bibitem{blanchet2008automated}Blanchet, B., Abadi, M. \& Fournet, C. Automated verification of selected equivalences for security protocols. {\em J Logic Algebr Progr}. \textbf{75}, 3-51 (2008)
\bibitem{ulitzsch2022post}Ulitzsch, V., Park, S., Marzougui, S. \& Seifert, J. A Post-Quantum Secure Subscription Concealed Identifier for 6G. {\em Proceedings Of The 15th ACM Conference On Security And Privacy In Wireless And Mobile Networks}. pp. 157-168 (2022)
\bibitem{hojjati2020blockchain}Hojjati, M., Shafieinejad, A. \& Yanikomeroglu, H. A blockchain-based authentication and key agreement (AKA) protocol for 5g networks. {\em IEEE Access}. \textbf{8} pp. 216461-216476 (2020)

\bibitem{regulation}Regulation, P. Regulation (EU) 2016/679 of the European Parliament and of the Council. {\em Regulation (eu)}. \textbf{679} pp. 2016 (2016)



\bibitem{chlosta20215g}Chlosta, M., Rupprecht, D., Pöpper, C. \& Holz, T. 5G SUCI-catchers: still catching them all?. {\em Proceedings Of The 14th ACM Conference On Security And Privacy In Wireless And Mobile Networks}. pp. 359-364 (2021)

\bibitem{eCK}LaMacchia, B., Lauter, K. \& Mityagin, A. Stronger security of authenticated key exchange. {\em International Conference On Provable Security}. pp. 1-16 (2007)

\bibitem{KEMTLS}Schwabe, P., Stebila, D. \& Wiggers, T. Post-quantum TLS without handshake signatures. {\em Proceedings Of The 2020 ACM SIGSAC Conference On Computer And Communications Security}. pp. 1461-1480 (2020)
\end{thebibliography}
%

\end{document}